\documentclass[aps,prb,reprint,shortbibliography,superscriptaddress]{revtex4-1}
\usepackage{ragged2e}
\usepackage{setspace}
\usepackage{amsmath}
\usepackage{graphicx}
\usepackage[nearskip,margin = 0pt]{subfig}
\usepackage{subfig}

\usepackage{verbatim}
\usepackage{amsfonts}
\usepackage{amssymb}
\usepackage{epstopdf} 
\usepackage{xcolor} 
\DeclareGraphicsExtensions{.pdf,.eps,.png,.jpg,.mps} 
\usepackage{hyperref}
\hypersetup{
    colorlinks=true,
    linkcolor=blue,
    citecolor=blue,
    filecolor=blue,      
    urlcolor=blue,
}
\begin{document}

\title{High-Q AlN microresonators for nonlinear near-infrared and near-visible photonics}

\author{Yulei Ding}
\affiliation{State Key Laboratory of Precision Measurement Technology and Instruments, Department of Precision Instruments, Tsinghua University, Beijing 100084, China}

\author{Yuming Huang}%
\affiliation{State Key Laboratory of Precision Measurement Technology and Instruments, Department of Precision Instruments, Tsinghua University, Beijing 100084, China}

\author{Zhongdong Yin}%
\affiliation{State Key Laboratory of Precision Measurement Technology and Instruments, Department of Precision Instruments, Tsinghua University, Beijing 100084, China}

\author{Yifei Wang}%
\affiliation{State Key Laboratory of Precision Measurement Technology and Instruments, Department of Precision Instruments, Tsinghua University, Beijing 100084, China}

\author{Kewei Liu}%
\affiliation{State Key Laboratory of Precision Measurement Technology and Instruments, Department of Precision Instruments, Tsinghua University, Beijing 100084, China}

\author{Yanan Guo}%
\affiliation{College of Materials Science and Opto-Electronic Technology, University of Chinese Academy of Sciences, Beijing 100049, China}
\affiliation{Research and Development Center for Wide Bandgap Semiconductors, Institute of Semiconductors, Chinese Academy of Sciences, Beijing 100083, China}

\author{Liang Zhang}%
\affiliation{College of Materials Science and Opto-Electronic Technology, University of Chinese Academy of Sciences, Beijing 100049, China}
\affiliation{Research and Development Center for Wide Bandgap Semiconductors, Institute of Semiconductors, Chinese Academy of Sciences, Beijing 100083, China}

\author{Qi Zhang}%
\affiliation{College of Materials Science and Opto-Electronic Technology, University of Chinese Academy of Sciences, Beijing 100049, China}
\affiliation{Research and Development Center for Wide Bandgap Semiconductors, Institute of Semiconductors, Chinese Academy of Sciences, Beijing 100083, China}

\author{Jianchang Yan}%
\affiliation{College of Materials Science and Opto-Electronic Technology, University of Chinese Academy of Sciences, Beijing 100049, China}
\affiliation{Research and Development Center for Wide Bandgap Semiconductors, Institute of Semiconductors, Chinese Academy of Sciences, Beijing 100083, China}

\author{Junxi Wang}%
\affiliation{College of Materials Science and Opto-Electronic Technology, University of Chinese Academy of Sciences, Beijing 100049, China}
\affiliation{Research and Development Center for Wide Bandgap Semiconductors, Institute of Semiconductors, Chinese Academy of Sciences, Beijing 100083, China}

\author{Changxi Yang}%
\affiliation{State Key Laboratory of Precision Measurement Technology and Instruments, Department of Precision Instruments, Tsinghua University, Beijing 100084, China}

\author{Chengying Bao}
\email{cbao@tsinghua.edu.cn}
\affiliation{State Key Laboratory of Precision Measurement Technology and Instruments, Department of Precision Instruments, Tsinghua University, Beijing 100084, China}

\maketitle

\newcommand{\ts}{\textsuperscript}

\newcommand{\tsb}{\textsubscript}

{\bf High Q-factors of microresonators are crucial for nonlinear integrated photonics, as many nonlinear dynamics have quadratic or even cubic dependence on Q-factors. The unique material properties make AlN microresonators invaluable for microcomb generation, Raman lasing and visible integrated photonics. However, the loss level of AlN falls behind other integrated platforms. By optimizing the fabrication, we demonstrate record Q-factors of 5.4$\times$10$^6$ and 2.2$\times$10$^6$ for AlN microresonators in the near-infrared and near-visible, respectively. Polarized-mode-interaction was used to create anomalous dispersion to support bright AlN Dirac solitons. Measurement of polarization-dependent spectra reveals the polarization hybridization of the Dirac soliton. In a microresonator with normal dispersion, Raman assisted four-wave-mixing (RFWM) was observed to initiate platicon formation, adding an approach to generate normal dispersion microcombs. A design of width-varying waveguides was used to ensure both efficient coupling and high Q-factor for racetrack microresonators at 780 nm. The microresonator was pumped to generate near-visble Raman laser at 820 nm with a fundamental linewidth narrower than 220 Hz. Our work unlocks new opportunities for integrated AlN photonics by improving Q-factors and uncovering nonlinear dynamics in AlN microresonators. }

\noindent \textbf{Introduction.} Reducing propagation loss in waveguides is fundamental to advancing photonic integrated circuits (PICs), as it directly impacts applications spanning from sensing, quantum photonics to optical frequency combs and low noise lasers \cite{Kippenberg_Science2018Review,Morandotti_NP2019quantum,yang2024efficient,Yang_Nature2017exceptional,Bao_NC2024integrated,Bowers_NP2021hertz}. The loss of CMOS-ready Si$_3$N$_4$ waveguides has been reduced to below 0.1 dB/m, corresponding to a Q-factor over 10$^8$ for microresonators \cite{Bowers_NP2021hertz,Blumenthal_NC2021422}. The success of PICs is inseparable from the contributions of various material platforms. For instance, lithium niobate (LN) or lithium tantalate (LT) can enable efficient and fast electro-optical modulation in PICs \cite{Loncar_Nature2018integrated,Boes_Science2023lithium,Kippenberg_Nature2024lithium}. Aluminum nitride (AlN) can further be used for piezoelectric modulation \cite{Bhave_AOP2024piezoelectric,Tang_AOP2023aluminum}. Unlike amorphous Si$_3$N$_4$, crystalline thin-film AlN exhibits strong nonlinear photon-phonon interactions, making it suitable for on-chip stimulated Raman and Brillouin relevant applications \cite{SunCZ_Optica2016,Bao_OL2022_Raman,Bao_PRL2025self,Li_Optica2019electromechanical}. Moreover, AlN features an ultrabroad transparency window with a bandgap as large as 6.2 eV, which holds great potential for visible and ultraviolet photonics \cite{Tang_AOP2023aluminum}. AlN has been used to attain high-Q microresonators \cite{Tang_Optica2018ultra,Mi_APL2021demonstration} and supercontinuum \cite{Tang_NC2019beyond} in the ultraviolet band. The large bandgap also means the material dispersion is less impacted by the absorption band, making it easier to engineer waveguide dispersion in the short wavelength. Due to this flexibility, octave spanning soliton microcombs have been demonstrated in AlN microresonators \cite{Tang_NC2021aluminum,Guo_PR2021directly}. With these octave spanning microcombs, $\chi^{(2)}$ nonlinearity of AlN could enable f-2f self-reference stabilization \cite{Tang_NC2021aluminum}. The  $\chi^{(2)}$ nonlinearity of AlN is intermediate, making it well-suited for investigating Pockels solitons or microcavity simultons reinforced by both $\chi^{(2)}$ and $\chi^{(3)}$ nonlinearities, as excessively strong $\chi^{(2)}$ nonlinearity can inhibit their formation \cite{Tang_NP2021pockels,Bao_PRL2024theoretical}.  Considering its potential applications, efforts have been devoted to optimizing the fabrication of AlN microresonators in the last decade \cite{stegmaier2014aluminum, Tang2015AlN,Liu2017aluminum,sun2019ultrahigh,Tang2019OPO,HUST2020photolithography,wang2021efficient,Tang_NC2021aluminum,Bao_OL2022_Raman,TangOL2025aluminum,zhang20253d}. However, the highest Q-factor has been limited to 3.7$\times$10$^6$ (a loss of 10 dB/m) \cite{Bao_OL2022_Raman,Tang_AOP2023aluminum}, which falls far behind the Si$_3$N$_4$ and LN platforms. Moreover, nonlinear dynamics in AlN microresonators has not been thoroughly studied. These hinder widespread use of nonlinear AlN photonics.

\begin{figure*}[t!]

\centering
{\includegraphics[width=\linewidth]{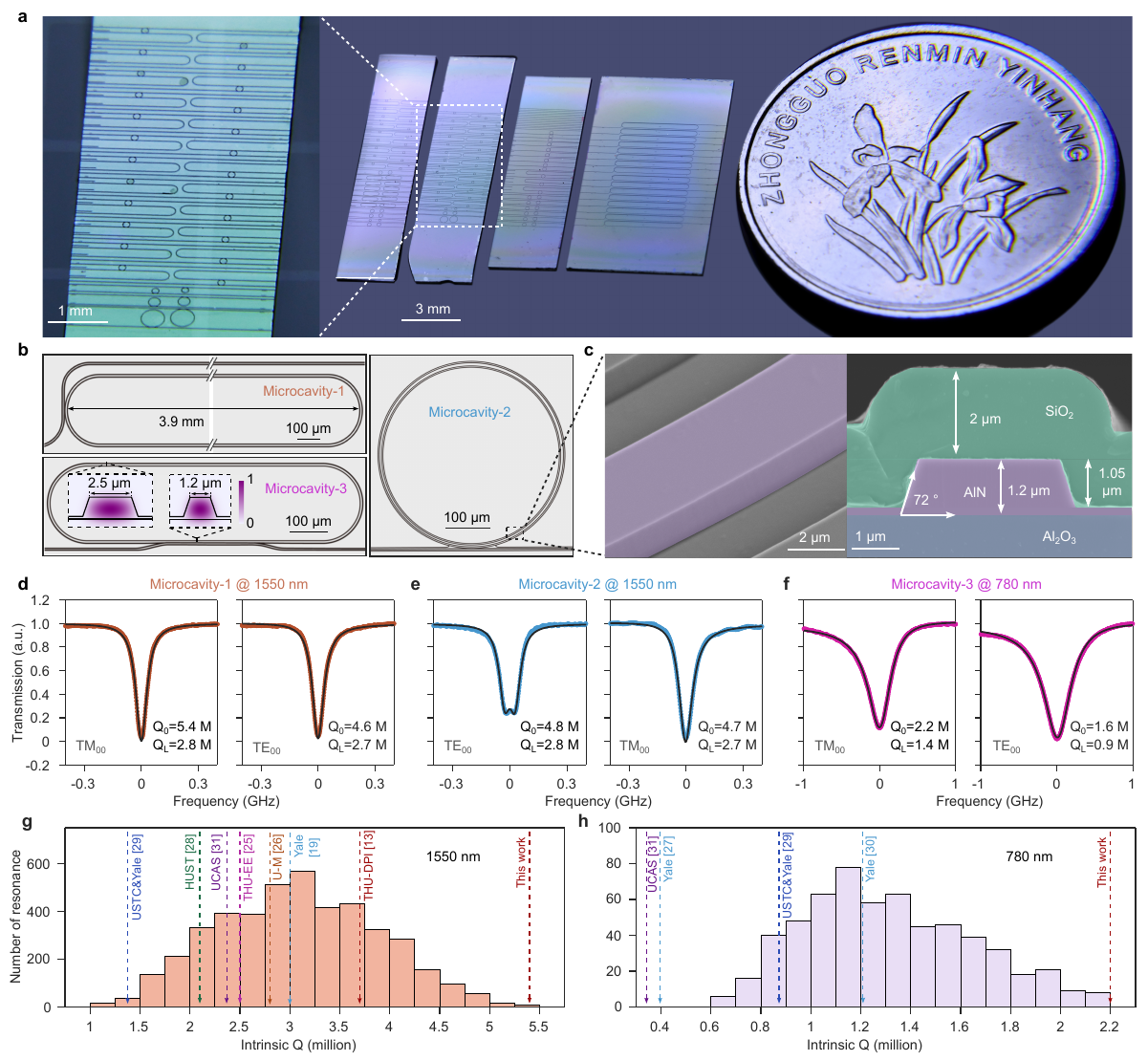}}
\captionsetup{singlelinecheck=off, justification = RaggedRight}
\caption{{\bf AlN photonic chips and Q-factor measurement.} 
\textbf{a,} Photograph of high-Q AlN-on-sapphire photonic chips.  
\textbf{b,} Optical microscope images of AlN microresonators. Microcavities-1, 2 are racetrack and microring designs for 1550 nm microresonators, while Microresonator-3 is a 780 nm microresonator. The inset shows the simulated mode profiles of the width-varying waveguide design for 780 nm microresonators.  
\textbf{c,} SEM images of the AlN waveguides. 
\textbf{d-f,} Measured high-Q resonance transmissions and their fits (black curve). Intrinsic and loaded Q-factors are labeled (assuming under-coupling). 
\textbf{g, h,} Distribution of intrinsic Q-factors measured for multiple microresonators in the 1550 nm and 780 nm band. The highest Q-factors of our samples are labeled to compare with other reports.} 
\label{fig1}
\end{figure*}

\begin{figure*}[t!]
\centering
{\includegraphics[width=\linewidth]{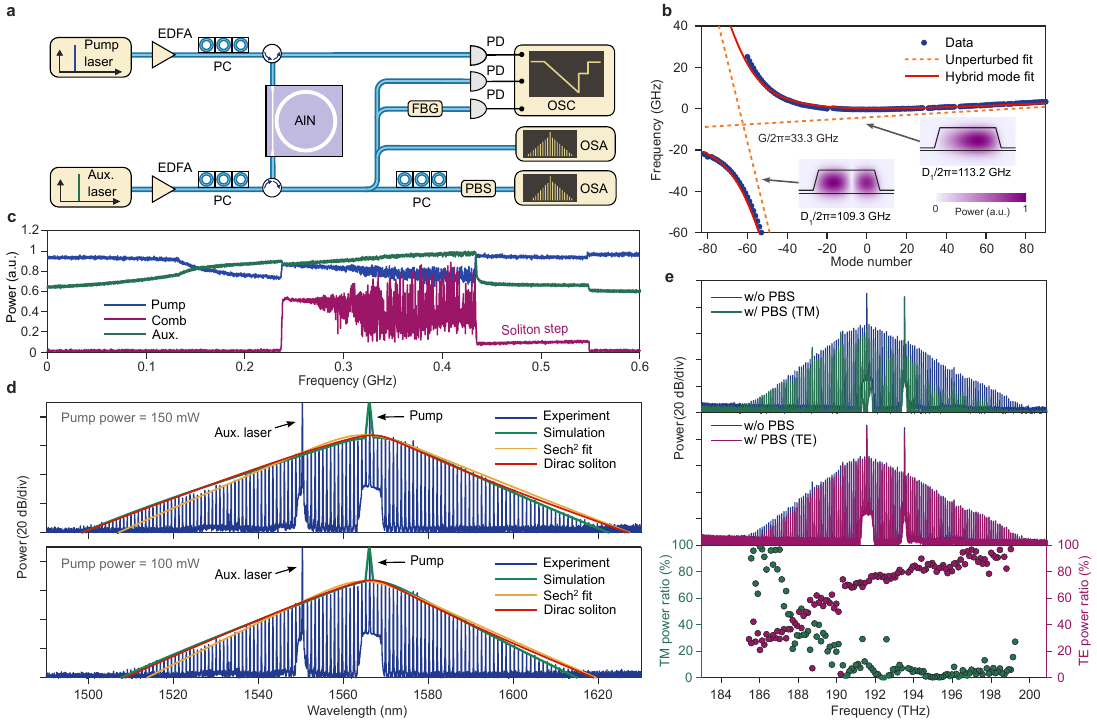}}
\captionsetup{singlelinecheck=off, justification = RaggedRight}
\caption{{\bf Polarization hybridized Dirac soliton microcomb in a high-Q AlN microresonator.} 
\textbf{a,} Experimental setup for soliton generation and polarization state characterization. EDFA, Erbium-doped fiber amplifier; PC, polarization controller; PBS, polarization beam splitter; FBG, fiber Bragg grating; PD, photodetector; OSA, optical
spectrum analyzer; OSC, oscilloscope. 
\textbf{b,} Measured hybridized mode frequencies (blue points) and the theoretical hybridization fit. Interactions between $\rm TE_{00}$ and $\rm TM_{10}$ modes creates anomalous dispersion (see inset for the unperturbed mode profiles).
\textbf{c,} Measured transmitted pump power, comb power and auxilliary pump power when tuning the pump laser to trigger the soliton step. 
\textbf{d,} Measured and simulated microcomb spectra using a pump power of 150 mW and 100 mW, respectively. The spectra deviate from a Sech$^2$ fit, but can be fitted by the Dirac soliton theory \cite{Vahala_LSA2020dirac}. 
\textbf{e,} Optical spectra measured when adjusting the PC to minimize or maximize the high frequency spectral part after the PBS. The bottom panel shows the polarization content across the spectrum.}
\label{fig2}
\end{figure*}

\begin{figure*}[t]
\centering
{\includegraphics[width=\linewidth]{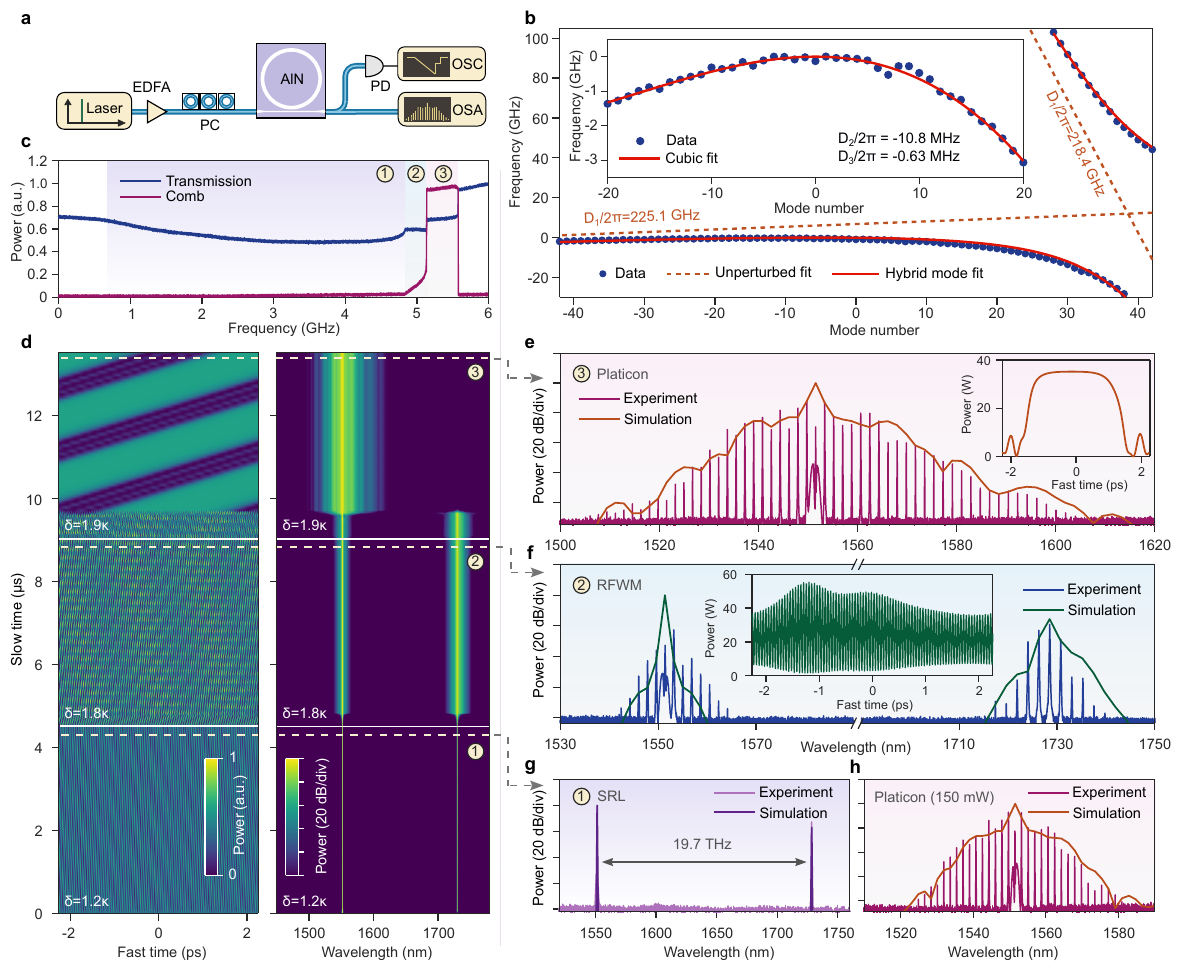}}
\captionsetup{singlelinecheck=off, justification = RaggedRight} 
\caption{{\bf RFWM initiated platicon microcomb in a normal dispersion AlN microresonator.} \textbf{a,} Experimental setup for platicon microcomb generation.
\textbf{b,} Measured $D_{\rm int}$ that is impacted by mode-interactions between TE${_{00}}$ and TM${_{10}}$ modes. The inset shows the zoomed-in view around the pumped mode, with the red curve being a cubic fit.
\textbf{c,} Measured transmitted and comb powers when scanning the pump laser across the resonance.
\textbf{d,} Simulated temporal and spectral dynamics of RFWM platicon formation, when the pump detuning is switched from 1.2$\kappa$ to 1.8$\kappa$ and then to 1.9$\kappa$ ($\kappa$ is the resonance linewidth).
\textbf{e-g,} Measured optical spectra of the platicon, RFWM, and SRL states under different detunings, corresponding to the three states labeled in panels c, d. Simulations are in agreement with the measurement and the insets show the simulated temporal waveforms.
\textbf{h,} Measured platicon microcomb generated with a lower pump power of 150 mW.}
\label{fig3}
\end{figure*}

\begin{figure*}[t]
\centering
\captionsetup{singlelinecheck=no, justification = RaggedRight}
\includegraphics[width=\linewidth]{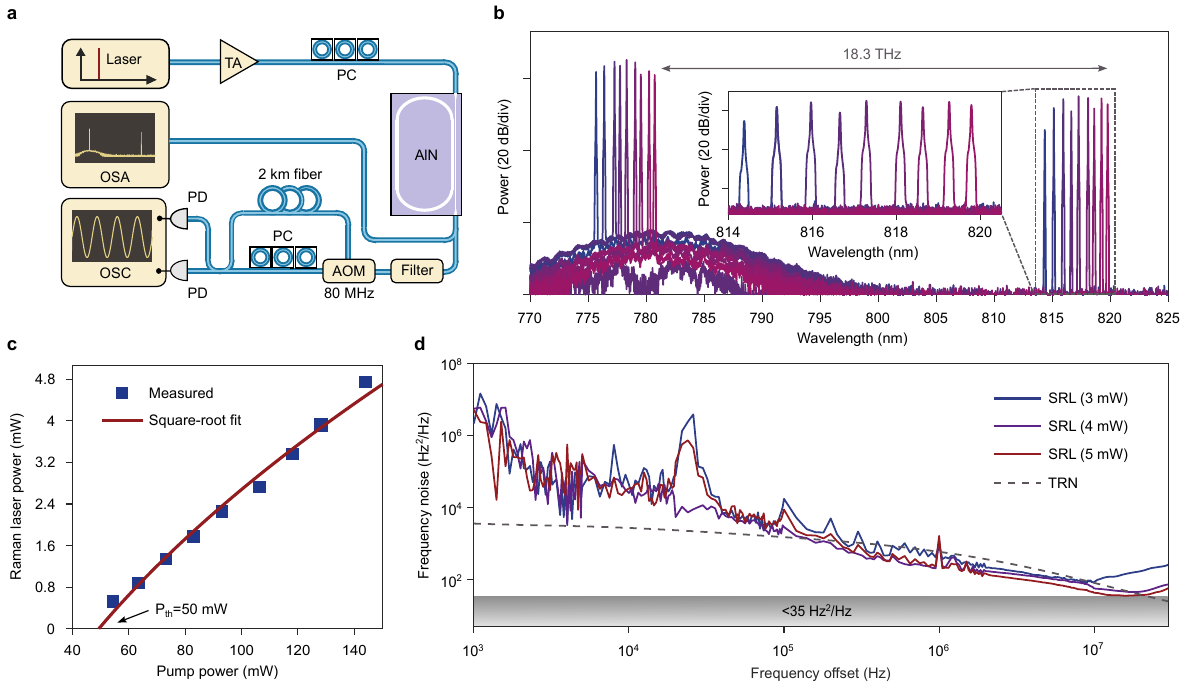}
\captionsetup{singlelinecheck=off, justification = RaggedRight}
\caption{{\bf Near-visible SRL in an AlN microresonator.} \textbf{a,} Experimental setup for near-visible SRL generation and its frequency noise measurement. 
TA, tapered amplifier. AOM, acousto-optical modulator. 
\textbf{b,}  Optical spectra of the SRLs using different pump wavelengths. The inset shows the zoomed-in view of the SRL spectra.
\textbf{c,} Measured on-chip SRL powers under different pump powers, exhibiting a square-root scaling and a threshold of 50 mW. 
\textbf{d,} Frequency noise of SRLs under different output powers, showing a fundamental noise lower than 35 Hz$^2$/Hz.}
\label{fig4}
\end{figure*}

In this work, we report single-crystalline AlN-on-sapphire microresonators with record Q-factors of 5.4$\times$10$^6$ in the near-infrared (1550 nm) and 2.2$\times$10$^6$ in the near-visible (780 nm). High-quality AlN film deposition and systematic fabrication optimization lead to the improved Q-factor. 
The microresonators were pumped to generate bright Dirac solitons \cite{Vahala_LSA2020dirac} and Raman assisted four-wave-mixing (RFWM) initiated dark pulses (platicons) \cite{Weiner_NP2015mode,Yang_PRA2020generation} in crystalline microresonators for the first time, to our knowledge. Although the fundamental mode of the AlN waveguide intrinsically features normal dispersion \cite{Bao_OL2021self,Bao_PRL2025self}, interactions between two polarized spatial modes were harnessed to create anomalous dispersion to support bright solitons \cite{Kippenberg_NP2014,Haus_IEEE2002coupled,Kippenberg_Science2018Review,Vahala_NC2017Visible,Vahala_LSA2020dirac,Kippenberg_NC2018photonic}. Measurements of the polarization-dependent spectra reveal polarization hybridization of the so-called bright Dirac soliton microcomb \cite{Vahala_LSA2020dirac}. We further observed the RFWM assisted platicon generation in a normal dispersion AlN microresonator. Such a generation dynamics was numerically predicted in our previous work \cite{Yang_PRA2020generation}, but has not been observed yet. In the near-visible band, a width-varying design was used to attain high-Q microresonator with efficient coupling. The high-Q factor further enables the generation of low noise stimulated Raman lasers (SRLs) with a fundamental linewidth narrower than 220 Hz at 820 nm by pumping at 780 nm. This is the second near-visible SRL in an integrated  microresonator after a diamond microresonator SRL \cite{Loncar2018Diamond780SRL}, to our knowledge. Our SRLs have the narrowest fundamental linewidth for near-visible, on-chip SRLs owing to the high Q-factor. 
The low loss AlN platform brings new opportunities for PICs, and our experiment adds insights to microcomb dynamics (including Dirac soliton and RFWM platicon dynamics). 

\noindent \textbf{AlN chips and Q-factor measurement.} The AlN microresonators (Fig. \ref{fig1}a) were fabricated using high-quality single-crystalline AlN films epitaxially grown by metal–organic chemical vapor deposition (MOCVD) on sapphire substrates (see surface roughness and X-ray diffraction measurements in Supplementary Sec. 1) \cite{Yan2011improved}. 
The fabrication process followed the procedure established in our previous work \cite{Bao_OL2021self, Bao_OL2022_Raman} (Methods). 

Both microring and racetrack resonators were fabricated (Fig. \ref{fig1}b). For example, Microcavity-1 (2) has a free spectral range (FSR) of 17 (113) GHz and is coupled by a bus waveguide with a width of 1.1 (1.3) $\mu$m and a coupling gap of 500 (650) nm.
The 1550 nm microresonators have a uniform waveguide width of 3.5 $\mu$m.  
The highest intrinsic Q-factor measured in the samples is 5.4$\times$10$^6$ (Fig. \ref{fig1}d), which was observed for a TM$_{00}$ mode in Microcavity-1. The highest Q-factor for TE$_{00}$ modes of this microresonator is 4.6$\times$10$^6$. For Microcavity-2, the highest Q-factor of TE$_{00}$ and TM$_{00}$ modes were 4.8$\times$10$^6$ and 4.7$\times$10$^6$ (Fig. \ref{fig1}e), respectively. 

Considering the tighter mode confinement for near-visible wavelengths, we varied the waveguide width along the racetrack for the 780 nm microresonator (Fig. \ref{fig1}b). In the coupling region, we used a relatively narrow 1.2 $\mu$m waveguide to enhance the external coupling rate. The line-coupling region has a length of 100 $\mu$m and the gap between the microresonator and the coupling waveguide is 600 nm. A wider 2.5 $\mu$m waveguides was used in the remaining region (about 70\% of the microresonator) to improve the Q-factor. Similar varying width design was also adopted in Si$_3$N$_4$ microresonators for visible photonics \cite{Lipson_NP2023widely}, but the Q-factor was not higher than 0.4$\times$10$^5$. We achieved the highest Q-factor of 2.2$\times$10$^6$ for the 780 nm AlN microresonators (TM$_{00}$ mode in Fig. \ref{fig1}f). For TE modes, the highest Q-factor was 1.6$\times$10$^6$. 


We characterized the intrinsic Q-factors from 1530 to 1570 nm for 8 racetrack resonators and from 1510 to 1590 nm for 4 microring resonators. We show the distribution of Q-factor measured for about 4400 resonances in Fig. \ref{fig1}g. The most probable Q-factor, which is also widely used to evaluate microresonators \cite{Kippenberg_NC2021high,VTC_OE2019high}, is 3.1$\times$10$^6$ for our samples. Since we focus on the possibly lowest loss for the AlN platform, we mainly use the highest Q-factor to evaluate our samples. The distribution of the intrinsic Q-factors for the near-visible microresonators (measured from 765 nm to 781 nm for 4 micresonators) is plotted in Fig. \ref{fig1}h, and the most probable Q-factor is 1.1 $\times$10$^6$. 

The 5.4$\times$10$^6$ and 2.2$\times$10$^6$ Q-factors constitute records for AlN microresonators in the 1550 nm band and 780 nm band, respectively (see other high-Q reports labeled in Figs. \ref{fig1}g, h). 
No evident dependence of Q-factor on polarization and resonator geometry (racetrack or microring) was observed (Supplementary Sec. 2).

\noindent \textbf{Hybridly polarized solitons in AlN microresonator.} We pumped Microcavity-2 with an FSR of 113 GHz to generate bright solitons in the telecom band (Fig. \ref{fig2}a).  
The fundamental TE$_{\rm 00}$ mode of the 3.5 $\mu$m wide AlN waveguide exhibits normal dispersion, and we pumped the anomalous dispersion TE$_{\rm 10}$ mode to generate bright solitons in our previous work \cite{Bao_OL2021self,Bao_PR2023mitigating}. Conversely, we pumped a hybridized fundamental mode to generate bright solitons in this work. Specifically, spatial-mode-interactions between the TE$_{\rm 00}$ and TM$_{\rm 10}$ modes was leveraged to attain anomalous dispersion \cite{Haus_IEEE2002coupled}. Similar scheme has been used to create anomalous dispersion to support bright solitons in the silica \cite{Vahala_NC2017Visible,Vahala_LSA2020dirac} and Si$_3$N$_4$ platforms \cite{Kippenberg_NC2018photonic}. Indeed, interactions between modes belonging to coupled Si$_3$N$_4$ microresonators have also been used to create anomalous dispersion for microcomb generation \cite{Qi_NC2017dispersion,Vahala_NP2023soliton}. Here, we adopted this approach in the AlN platform for the first time, to our knowledge. 

Dispersion of the microresonator is evaluated as $D_{\rm int}$=$\omega(\mu)-\omega(0)-\mu D_1$, where $\omega(\mu)$ is the resonance frequency for mode-$\mu$ and $D_1/2\pi$ is the FSR of the microresonator. $D_{\rm int}$/2$\pi$ of the pumped hybridized mode was measured by sweeping a tunable laser to map out the resonances (Fig. \ref{fig2}b). Based on the coupled-mode theory \cite{Haus_IEEE2002coupled, Vahala_Optica2016DW}, we fitted the measured resonances and derived a coupling rate between the two modes as $G$/2$\pi$=33.3 GHz (Methods). It is essential for $G$/2$\pi$ to be much larger than FSR difference between the two interacting modes to create broadband anomalous dispersion. This FSR difference between the interacting TE$_{\rm 00}$ and TM$_{\rm 10}$ modes is fitted to be 3.9 GHz, satisfying the requirement. 

The dual-pump scheme was used to mitigate thermal instability during soliton generation in the AlN microresonator \cite{Zhou_LSA2019soliton,Del'Haye_Optica2019sub,Bao_OL2021self} (Fig. \ref{fig2}a). Other techniques including the engineered fast pump laser sweep can also be used to stabilize solitons in AlN microresonators \cite{Papp_PRL2018,Bao_PR2023mitigating,Tang_NC2021aluminum}. By setting the auxiliary pump appropriately at 1550.3 nm, the soliton step can be extended when pumping at 1566.1 nm (Fig. \ref{fig2}c). The pumped mode have a Q-factor of 3.5$\times$10$^6$ (Supplementary Sec. 2). Note that we did not pump the high-Q mode of 4.8$\times$10$^6$ located at 1572 nm, which is outside the gain bandwidth of the erbium-doped fiber amplifier (EDFA). We generated solitons with a relatively low on-chip pump power of 150 mW or 100 mW (Fig. \ref{fig2}d). This is also the lowest repetition rate for AlN microcombs. Indeed, the high Q-factor enables the lowest metrics of pump power times FSR for AlN microcombs (see Supplementary Table 1). 

The mode-interaction-induced dispersion is relatively large and contains large high-order mode dispersion (fitted second-order to fourth-order dispersion coefficients are $\beta_2$=$-$516 ${\rm ps^2/km}$, $\beta_3$=38.4 ${\rm ps^3/km}$, and $\beta_4$=$-$1.74 ${\rm ps^4/km}$). Hence, the generated microcomb is relatively narrow and asymmetric, which deviates from conventional sech$^2$-shape and can be fitted by a Dirac soliton spectrum (Supplementary Sec. 3) \cite{Vahala_LSA2020dirac}. When using the measured dispersion for simulation based on the generalized Lugiato-Lefever equation (LLE) \cite{Coen_OL2013modeling,Kippenberg_Science2018Review,Bao_OL2021self}, the simulated spectra are in excellent agreement with the measurement (Methods). Note that the FSR is smaller than the Raman gain bandwidth (114 GHz) and strong interplay with Raman gain in microcomb generation was observed when pumping other modes (Supplementary Sec. 4). Increasing the external coupling rate for the Raman gain band by appropriate design of the coupling waveguide can be adopted to minimize the Raman influence in microcomb generation \cite{Loncar_LSA2024octave}.

To verify the generated microcomb is supported by polarization-hybridized modes, we inserted a fiber-based polarization beam splitter (PBS) before the optical spectrum analyzer to measure the polarization-dependent spectra. In the experiment, we adjusted the polarization controller (PC) to maximize or minimize the power on the high frequency part (positive modes) of the spectra (Fig. \ref{fig2}e). Polarization contents in these two cases differ in the low frequency part (negative modes), suggesting the polarization state of the generated soliton varies across the spectrum. While the high frequency part power can decrease to a very low level, the low frequency part power did not decrease to a very low level in either case. It suggests positive modes have a relatively pure TE polarization, while negative modes are considerably hybridized (see the calculated polarization content in the bottom panel of Fig. \ref{fig2}e). This is consistent with measured resonance frequencies in Fig. \ref{fig2}b. The mode-resolved polarization measurement adds insight to Dirac soliton dynamics supported by hybridized modes.

\noindent \textbf{Platicon microcombs in AlN microresonators.} In addition to bright solitons, dark pulse or platicon microcombs are also important, as it can enhance the output comb power and enable microcomb generation in visible wavelengths featuring large normal dispersion \cite{Weiner_NP2015mode,Weiner_LPR2017microresonator}. Self-injection locked platicon microcombs have been demonstrated in AlN microresonators recently \cite{Bao_PRL2025self}; however, conventionally pumped AlN platicon microcombs have not been demonstrated, to our knowledge. 

Such a conventional pumping scheme (Fig. \ref{fig3}a) can enable broader comb generation, but usually needs assistance of local dispersion perturbation induced by mode-interactions \cite{Weiner_NP2015mode,Weiner_PRL2018}. RFWM was predicted to provide another route for the formation of platicons \cite{Yang_PRA2020generation}, but has not been observed yet. We observed this dynamics in one of our samples with an intrinsic Q-factor of 2.4$\times$10$^6$ (Supplementary Fig. S3) and an FSR of 225.1 GHz. The measured $D_{\rm int}$ for this microresonator is shown in Fig. \ref{fig3}b. Mode-interactions between TE$_{00}$ and TM$_{10}$ modes were also observed in this microresonator. However, we pumped at 1551.5 nm (mode-0 in Fig. \ref{fig3}b), which is not strongly impacted by the interactions. Therefore, microcomb generation in this microresonator is mainly governed by the platicon dynamics. 
At a pump power of 300 mW, the platicon can be generated by manually tuning the pump into the resonance, and the observed transmitted pump and comb powers are plotted in Fig. \ref{fig3}c. Three operation regimes were observed. These regimes were simulated and shown in Fig. \ref{fig3}d. When tuning the pump into the resonance, SRL was first observed (Figs. \ref{fig3}d, g). By further tuning into the resonance, frequencies were generated around the pump and the SRL via RFWM \cite{Yang_PRA2020generation} (Figs. \ref{fig3}d, f). This state finally transitioned into a platicon state by further increasing the pump wavelength (Figs. \ref{fig3}d, e). Meanwhile, comb lines in the Raman band annihilated. The measured spectra are in reasonable agreement with simulations (Figs. \ref{fig3}e-g). The stabilized microcomb is asymmetric with less lines located in the short wavelength side (Fig. \ref{fig3}e), which can be attributed to the higher dispersion for positive modes due to mode-interactions (inset of Fig. \ref{fig3}b). Besides the broadband mode-interaction, simulations also suggest localized mode-interaction induced resonance shift (about 0.13 GHz for the pumped mode-0) also helps with platicon formation (Supplementary Sec. 5). The RFWM platicon can also be generated with an on-chip pump power as low as 150 mW with a slightly narrower comb bandwidth (Fig. \ref{fig3}h). 

\noindent \textbf{Near-visible stimulated Raman lasers.} Besides microcomb, SRL is another important application of integrated microresonators. The large optical phonon frequency of AlN can shift the lasing frequency into spectral regions that are hard to access. Furthermore, high-Q resonators can help purify the pump wave via SRL, as it endows a narrower Schawlow-Townes linewidth \cite{Bao_OL2022_Raman}. 
Microresonator SRLs were usually pumped in the telecom band. On-chip near-visible SRL has only been demonstrated in a diamond microresonator \cite{Loncar2018Diamond780SRL}, to our knowledge. Frequency noise of the SRL was not characterized in that work, but could be relatively high considering its relatively low Q-factors (0.3$\times$10$^6$ at 750 nm) \cite{Loncar2018Diamond780SRL}. Here, we generate low noise SRLs around 820 nm by pumping the AlN microresonator at 780 nm with an FSR of 80 GHz.

In the experiment, we pumped TM$_{\rm 00}$ modes of the microresonator for SRL generation. The phonon frequency of TM modes is 18.3 THz, slightly lower than the 19.7 THz of TE modes \cite{Bao_OL2022_Raman,SunCZ_Optica2016}. SRLs were observed by pumping multiple modes of the microresonators (Fig. \ref{fig4}b). As a result, the SRL wavelength can be shifted by 6 nm (inset of Fig. \ref{fig4}b), limited by the bandwidth of the 780 nm semiconductor tapered amplifier. The output power was measured to increase with the pump power, and a square-root fit of the output power shows the threshold power is 50 mW (Fig. \ref{fig4}c). 

Then, we measured frequency noise of the SRL at 820 nm using the delayed self-heterodyne technique \cite{Bao_OL2022_Raman,Bowers_NP2021hertz} (Figs. \ref{fig4}a, d). Frequency noise for offset frequencies below 100 kHz can be impacted by the pump laser. Above 100 kHz, frequency noise of the SRLs is limited by the thermo-refractive noise (TRN), considering the small mode volume. The dashed line in Fig. \ref{fig4}d represents the simulated TRN \cite{Bao_PRL2025self,Kippenberg_PRA2019thermorefractive}. The lowest frequency noise measured in the experiment decreases with the SRL power (frequency noise rises at high offset frequencies due to the measurement noise floor). And, the fundamental linewidth should be narrower than 220 Hz for the 5 mW SRL. This is slightly larger than the 1750 nm AlN SRL \cite{Bao_OL2022_Raman,Bao_PRL2025self}, but consistutes the narrowest linewidth for on-chip SRL in the near-visible band, to our knowledge.

\noindent \textbf{Discussions.} In summary, we have improved record Q-factors of AlN microresonators to 5.4$\times$10$^6$ at 1550 nm and 2.2$\times$10$^6$ at 780 nm, respectively. Our measurement further suggests the absorption limited Q-factor (loss) of the AlN platform can be up to 8.0$\times$10$^6$ (4.6 dB/m) at 1550 nm and 3.0$\times$10$^6$ (24.5 dB/m) at 780 nm, see Supplementary Sec. 6 \cite{Blumenthal_NC2021422,Gao_NC2022_probing}. The demonstrated high-Q AlN platform was used for the generation of Dirac soliton, RFWM platicon microcomb and low-noise near-visible SRL. Our work reveals how mode-interactions and RFWM can add to the functionality of the AlN platform. Our work also experimentally verifies that RFWM is a viable route for the formation of platicon microcomb. Microresonators with varying-width designs provide a promising route towards visible AlN nonlinear photonics. 

High-Q AlN microcombs can be counter-pumped to generate counter-propagating solitons, which can be used to study soliton interaction physics and to build dual-comb sensing systems \cite{Bao_PRX2025rhythmic,Vahala_NP2017Counter,Vahala_NP2021quantum}. Counter-pumped AlN microresonators can also be used for stimulated Brillouin lasers and integrated optical gyroscope \cite{Vahala_NP2020earth}. Piezoelectric  or electro-optical response of AlN can be harnessed to achieve fully monolithic laser or microcomb modulation \cite{Tang_AOP2023aluminum,Bhave_AOP2024piezoelectric,TangOL2025aluminum}. The demonstrated AlN microresonators are not suitable for $\chi^{(2)}$ nonlinear photonics, as they were not designed for phase-matching. Intermodal phase-matching can be used to enable $\chi^{(2)}$ nonlinear photonics in AlN microresonators. High-Q factors and phase-matching can enable efficient generation of entangled photon pairs \cite{Tang_LSA2017parametric}, squeezed microcombs \cite{Yi_NC2021squeezed}, and microcavity simulton-based mid-infrared microcombs \cite{Tang_NP2021pockels,Bao_PRL2024theoretical}. The observed dynamics can also be harnessed to advance nonlinear photonic applications in other integrated crystalline platforms including GaN, LN, LT and SiC.

\vspace{3mm}
\noindent \textbf{Methods} 

{\small
\vspace{1 mm}
\noindent \textbf{Fabrication of AlN microresonators.} 
The AlN microresonators were fabricated using single-crystalline AlN films epitaxially grown on c-plane sapphire substrates by MOCVD. A two-step etching process was implemented to etch the AlN film. First, a 500-nm-thick Si$_3$N$_4$ hard mask layer was deposited on the AlN film via plasma-enhanced chemical vapor deposition (PECVD). Device patterns were defined by electron-beam lithography (EBL). A conductive polymer layer was spin-coated to effectively suppress charge accumulation during exposure. The patterns were subsequently transferred to the Si$_3$N$_4$ hard mask through reactive ion etching (RIE). Then, the AlN film was etched by inductively coupled plasma (ICP). After removing the hard mask, a 2 $\mu$m-thick-SiO$_2$ cladding layer was deposited by PECVD. Finally, the chips were annealed at 1050 $^\circ$C for 2 hours.

\vspace{1 mm}
\noindent \textbf{Hybridized mode resonances.} 
Based the coupled-mode theory \cite{Haus_IEEE2002coupled}, the resonances of hybridized microresonator modes can be written as, 
\begin{equation}
{\omega_\pm }\left( \mu  \right) = \frac{{{\omega _{\rm 1}}\left( \mu  \right) + {\omega _{\rm 2}}\left( \mu  \right)}}{2} \pm \sqrt {{G^2} + \frac{{{{\left[ {{\omega _{\rm 1}}\left( \mu  \right) - {\omega_{\rm 2}(\mu)}} \right]}^2}}}{4}},
\label{eq1}
\end{equation}
where $\omega_\pm(\mu)$ is the resonance for mode-$\mu$ in the two hybridized branches, $G$ is the coupling rate, and $\omega_{1(2)}(\mu)$ is the initial resonance for the two interacting mode families. 

For Fig. \ref{fig2}b, the two interacting modes are TE$_{00}$ and TM$_{10}$ modes, and the coupling rate is fitted to be $G/2\pi$=33.3 GHz. This coupling rate is fitted to be $G/2\pi$=43.0 GHz for TE$_{00}$ and TM$_{10}$ modes in Fig. \ref{fig3}b. 

\vspace{1 mm}
\noindent \textbf{Numerical simulations.} We used the generalized LLE to simulate microcomb generation dynamics. The simulation model is written as \cite{Bao_OL2021self},

\begin{equation}
\begin{aligned}
& \frac{{\partial {A}}}{{\partial T}}  = \left( { - \frac{\kappa }{2} - i\delta \omega  - \frac{{{i\beta_2}L}}{{2{T_{\rm R}}}}\frac{{{\partial ^2}}}{{\partial {\tau ^2}}}+\frac{{{\beta _3}L}}{{6{T_{\rm R}}}}\frac{{{\partial ^3}}}{{\partial {\tau ^3}}}+\frac{{{i\beta _4}L}}{{24{T_{\rm R}}}}\frac{{{\partial ^4}}}{{\partial {\tau ^4}}}} \right){A} \\ & + i\frac{{\gamma L}}{{{T_{\rm R}}}} {{{\left| A \right|}^2}} A   + i\frac{{{\gamma _{\rm R}}L}}{{{T_{\rm R}}}}\left[ {{A}\mathop \int \nolimits_{ - \infty }^\tau  {h_{\rm R}}\left( {\tau  - \tau '} \right){{\left| {{A}} \right|}^2}d\tau '} \right] +\sqrt{ {\frac{{\kappa _{\rm e}}P_{\rm in}}{T_R}}},
\label{eqAF}
\end{aligned}
\end{equation}
where $A$ is the envelope of the intracavity field, $T$ is slow time, $L$ is the microcavity length, $T_{\rm R}$ is the round trip time; $\kappa=\kappa_0+\kappa_{\rm e}$ is the total loss rate ($\kappa_0$ is the intrinsic loss rate and $\kappa_e$ is the external coupling loss rate), $\delta\omega$ is the pump–resonance frequency detuning, and  $\beta_{2}$, $\beta_{3}$, $\beta_{4}$ are the group velocity dispersion, third-order dispersion and fourth-order dispersion, respectively; $\gamma$ is the Kerr nonlinearity coefficient,  $\gamma_{\rm R}$ is Raman nonlinearity coefficient and $h_{\rm R}$ is Raman response function. The Raman action is calculated in the frequency domain with a Lorentzian gain spectrum centered at $-2\pi\times$19.7 THz having a bandwidth of $2\pi\times$114 GHz \cite{Bao_OL2021self}. 

The simulation parameters for Fig. \ref{fig2} include $\kappa$=2$\pi\times$98 MHz, $\kappa_{\rm e}$=2$\pi\times$44 MHz, $L$=1.26 mm, $T_{\rm R}$=8.8 ps, $\beta_2$=$-$516 ${\rm ps^2/km}$, $\beta_3$=38.4 ${\rm ps^3/km}$, and $\beta_4$=$-$1.74 ${\rm ps^4/km}$, $\gamma$=0.55 (Wm)$^{-1}$, $\gamma_{\rm R}$=0.0035 (Wm)$^{-1}$, $\delta\omega$ was chosen as 3.65$\kappa$ and 3$\kappa$ for the pump power of 150 mW and 100 mW, respectively. 
For Fig. \ref{fig3}, the used simulation parameters are $\kappa$=2$\pi\times$145 MHz, $\kappa_{\rm e}$=2$\pi\times$65 MHz, $L$=0.63 mm, $T_{\rm R}$=4.4 ps, $\beta_2$=240 ${\rm ps^2/km}$, $\beta_3$=10 ${\rm ps^3/km}$, and $\beta_4$=0 ${\rm ps^4/km}$, $\gamma$=0.55 (Wm)$^{-1}$, $\gamma_{\rm R}$=0.0035 (Wm)$^{-1}$, $P_{\rm in}$=200 mW (Fig. \ref{fig3}e) or 120 mW (Fig. \ref{fig3}h). The used pump powers were slightly lower than the experiment, which can be attributed to the exclusion of backscattering and wavelength-dependent Q-factors in the simulation (see also Supplementary Fig. S3). Note that we also included mode-interaction induced resonance shift for the pumped mode for simulations in Fig. \ref{fig3}. This resonance shift was added in the frequency domain by adding a phase shift per round trip $\Delta\phi$ = $-$2$\pi \Delta f T_R$, where $\Delta f$ is the resonance shift set as 0.13 GHz \cite{Weiner_PRL2018}. More discussions on the role of mode-interaction in RFWM platicon generation can be found in Supplementary Sec. 5. 

\vspace{1 mm}
\noindent \textbf{Data Availability.}
The data that supports the plots within this paper and other findings is available.

\vspace{1 mm}

\noindent \textbf{Code Availability}
The code that supports findings of this study are available from the corresponding author upon reasonable request.

\vspace{1mm}
\noindent\textbf{Acknowledgment.} This work is supported by the National Natural Science Foundation of China (62250071, 62175127, 62375150), by the National Key Research and Development Program
of China (2021YFB2801200, 2023YFB3211200), and by the Tsinghua-Toyota Joint Research Fund.

\vspace{1mm}
\noindent\textbf{Author contribution.} The AlN microresonators was fabricated by Y.D., Y.H. and K.L. Measurement and analysis was led by Y.D. with assitance from Y.H., Z.Y., Y.W. and C.Y. The AlN films were deposited by Y.G., J.Y., L.Z., Q.Z. and J.W. The project was supervised by C.B. 

\vspace{1mm}
\noindent \textbf{Competing Interests.} The authors declare no competing interests.

\bibliography{main.bib}

\end{document}